\documentclass[aps,pra,twocolumn,superscriptaddress]{revtex4-1}

\usepackage{graphicx}
\usepackage{amsmath}

\begin{document}

\title{Superbunching pseudothermal light}

\author{Yu Zhou}
\affiliation{MOE Key Laboratory for Nonequilibrium Synthesis and Modulation of Condensed Matter, Department of Applied Physics, Xi'an Jiaotong University, Xi'an 710049, China}

\author{Bin Bai}
\affiliation{Electronic Materials Research Laboratory, Key Laboratory of the Ministry of Education \& International Center for Dielectric Research, Xi'an Jiaotong University, Xi'an 710049, China}

\author{Huaibin Zheng}
\affiliation{MOE Key Laboratory for Nonequilibrium Synthesis and Modulation of Condensed Matter, Department of Applied Physics, Xi'an Jiaotong University, Xi'an 710049, China}
\affiliation{Electronic Materials Research Laboratory, Key Laboratory of the Ministry of Education \& International Center for Dielectric Research, Xi'an Jiaotong University, Xi'an 710049, China}

\author{Hui Chen}
\affiliation{Electronic Materials Research Laboratory, Key Laboratory of the Ministry of Education \& International Center for Dielectric Research, Xi'an Jiaotong University, Xi'an 710049, China}

\author{Jianbin Liu}
\email[]{liujianbin@xjtu.edu.cn}
\affiliation{Electronic Materials Research Laboratory, Key Laboratory of the Ministry of Education \& International Center for Dielectric Research, Xi'an Jiaotong University, Xi'an 710049, China}

\author{Fu-li Li}
\affiliation{MOE Key Laboratory for Nonequilibrium Synthesis and Modulation of Condensed Matter, Department of Applied Physics, Xi'an Jiaotong University, Xi'an 710049, China}

\author{Zhuo Xu}
\affiliation{Electronic Materials Research Laboratory, Key Laboratory of the Ministry of Education \& International Center for Dielectric Research, Xi'an Jiaotong University, Xi'an 710049, China}

\date{\today}

\begin{abstract}
A novel and simple superbunching pseudothermal light source is introduced based on common instruments such as laser, lens, pinhole and groundglass. $g^{(2)}(0)=3.66 \pm 0.02$ is observed in the suggested scheme by employing two rotating groundglass.  Quantum and classical theories are employed to interpret the observed superbunching effect. It is predicted that $g^{(2)}(0)$ can reach $2^N$ if $N$ rotating groundglass were employed. These results are helpful to understand the physics of superbunching. The proposed superbunching pseudothermal light may serve as a new type of light to study the second- and higher-order coherence of light and have potential application in improving the visibility of thermal light ghost imaging.
\end{abstract}

\maketitle

\section{Introduction}\label{introduction}

Two-photon bunching was first observed by Hanbury Brown and Twiss in 1956 \cite{hbt,hbt-1}, in which randomly emitted photons by thermal light source were found to have the tendency to come in bunches rather than randomly. Lots of attentions were drawn to this bunching effect shortly after it was reported. Some researchers repeated Hanbury Brown and Twiss's experiments and got negative results \cite{brannen,ferguson}. It was later understood that the negative results were due to the response time of the detection system is much longer than the coherence time of the measured light beams \cite{mandel-book}. Classical theory was first employed  to interpret the bunching effect \cite{forrester,forrester1,brown-1957,brown-1958}. Then quantum theory was also employed to interpret the same effect \cite{purcell,fano,glauber,glauber1}. It is now well accepted that the two-photon bunching effect of thermal light can be explained by both quantum and classical theories \cite{glauber,glauber1,sudarshan}. However, the full quantum explanation of the bunching effect given by Glauber greatly deepens our understanding of optical coherence. The Hanbury Brown and Twiss's experiments \cite{hbt,hbt-1} and Glauber's quantum optical coherence theory \cite{glauber,glauber1} are usually thought as the cornerstones of modern quantum optics \cite{glauber2}.

The experimental setup employed by Hanbury Brown and Twiss \cite{hbt,hbt-1}, which is known as Hanbury Brown-Twiss (HBT) interferometer, plays an important role in measuring the second-order coherence of light and photon statistics in quantum optics \cite{scully-book,mandel-book}. The second-order coherence of light can be described by the normalized second-order coherence function introduced by Glauber \cite{glauber,glauber1}. For a light beam in a HBT interferometer, two-photon bunching is defined as $g^{(2)}(0)>g^{(2)}(\tau)$ ($\tau \neq 0$), where $g^{(2)}(\tau)$ is the normalized second-order coherence function and $\tau$ is the time difference between two photon detection events within a two-photon coincidence count. On the contrary, antibunching is defined as  $g^{(2)}(0)<g^{(2)}(\tau)$ ($\tau \neq 0$), which is usually thought as a nonclassical effect \cite{mandel-book}. It is well-known that $g^{(2)}(0)$ equals 2 for thermal light. For the bunched light which satisfies $g^{(2)}(0)>2$, the word \textit{superbunching} is usually employed \cite{ficek-book}.

Two-photon superbunching is usually introduced by nonlinear interaction between light and atoms \cite{lipeles,kaul,kocher,akhmanov,swain,auffeves,hoi,grujic,bhatti}, quantum dots \cite{albert,jahnke,redlich}, or nonlinear medium \cite{klyshko,burnham,alley,ou,chekhova,bromberg}, \textit{etc}.. The efficiency of generating two-photon bunching effect with nonlinear interaction is usually very low and high precision is always required in adjusting the experimental setup \cite{lipeles,kaul,kocher,akhmanov,swain,auffeves,hoi,grujic,bhatti,albert,jahnke,redlich,klyshko,burnham,alley,ou,chekhova,bromberg}. Fortunately, nonlinear interaction is not necessary for generating two-photon superbunching. For instance, Hong and one of the present authors \textit{et al.} observed $g^{(2)}(0)=2.4\pm 0.1$ in a linear system via multiple two-photon path interference \cite{hong}. They further predicted that higher value of $g^{(2)}(0)$ could be reached by adding more pathes. However, it is a big experimental challenge if more pathes were added in their scheme \cite{hong}. In this paper, we will introduce a different and much simpler scheme for superbunching with classical light in a linear system, which is called superbunching pseudothermal light. The corresponding quantum and classical interpretations of the superbunching effect in our scheme are of great importance to understand the physics of superbunching. The proposed superbunching pseudothermal light source may find possible applications in the second- and higher-order interference of light \cite{scully-book,mandel-book} and high-visibility ghost imaging with classical light  \cite{shih-book}.

The paper is organized as follows. In Sect. \ref{theory}, we will introduce the superbunching pseudothermal light source and employ two-photon interference theory to calculate the second-order coherence function. The experimental setup with two rotating groundglass is employed to observe superbunching in Sect. \ref{experiments}. The discussions about the physics of superbunching and an alternative scheme for superbunching pseudothermal light source are in Sect. \ref{discussions}. Section \ref{conclusion} summarizes our conclusions.

\section{Theory}\label{theory}

\subsection{Superbunching pseudothermal light with $N$ rotating groundglass}
\begin{figure}[htb]
\centering
\includegraphics[width=62mm]{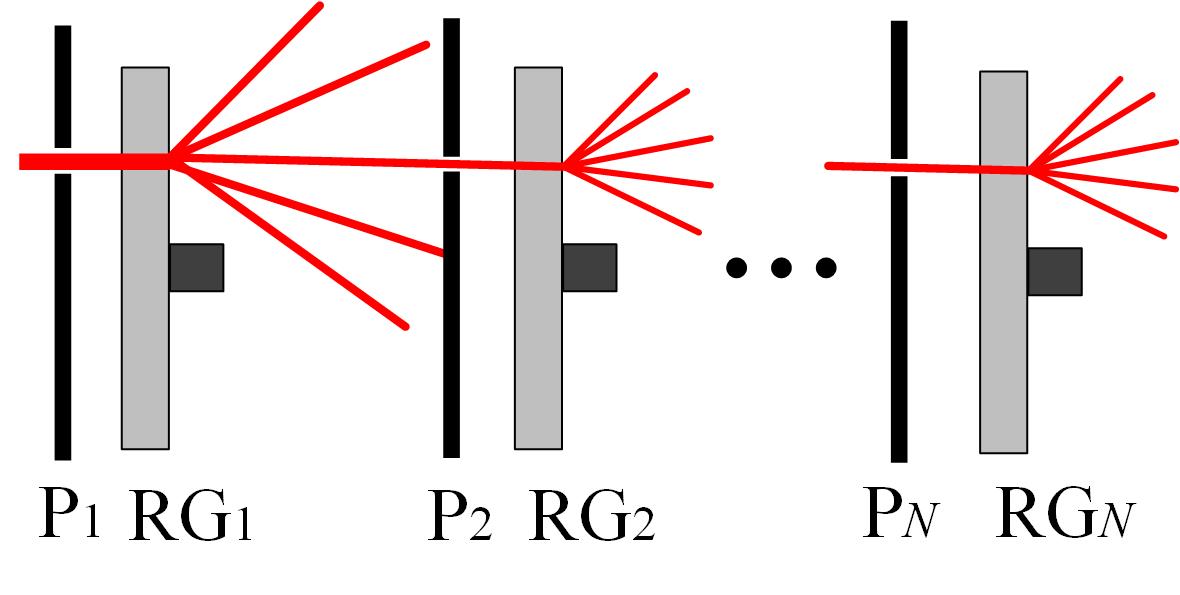}%
\caption{(Color online) Superbunching pseudothermal light source. P$_j$: the $j$th pinhole. RG$_j$: the $j$th rotating groundglass. The pinhole before RG$_j$ is employed to filter out the light beam within one coherence area of pseudothermal light generated by RG$_{j-1}$ ($j=2$, 3,..., $N$).}\label{1}
\end{figure}

The superbunching pseudothermal light source is shown in Fig. \ref{1}, where P$_j$ and RG$_j$ are the $j$th pinhole and rotating groundglass, respectively ($j=1$, 2, ..., $N$). A coherent light beam is incident to RG$_1$ after passing through P$_1$. The scattered light is then filtered out by P$_2$. The filtered light beam is within the same transverse coherence area of pseudothermal light generated by RG$_1$. The light incident on RG$_2$ is coherent and there will be interference pattern after RG$_2$. Another pinhole and RG can be put after RG$_2$ in the same manner and the process can be repeated for $N$ (positive integer) RGs.

If a single-mode continuous-wave laser light beam is employed as the input before P$_1$ in Fig. \ref{1}, the scattered light after RG$_1$ is pseudothermal light \cite{martienssen}, which has been applied extensively in thermal light ghost imaging \cite{gatti-2004,chen-2004,valencia,cai}, the second- and higher-order interference of thermal light \cite{mandel-1983,liu-2009,liu-2013,liu-epl}. The photons in light beam after $N$ ($N \geq 2$) RGs will be superbunched, which means the normalized second-order coherence function, $g^{(2)}(0)$, will exceed 2.

The HBT interferometer \cite{hbt,hbt-1} is employed in the scheme shown in Fig. \ref{1} to measure the second-order coherence function. When there is only one RG in the scheme, there are two different alternatives for two photons in pseudothermal light to trigger a two-photon coincidence count. One is photon a (short for photon at position a) goes to D$_1$ (short for detector 1) and photon b goes to D$_2$. The other one is photon a goes to D$_2$ and photon b goes to D$_1$. If these two different alternatives are indistinguishable, the second-order coherence function is \cite{purcell,fano,feynman-qed,feynman-path,shih-book}
\begin{eqnarray}\label{g2-one}
G^{(2)}_1(\mathbf{r}_1,t_1;\mathbf{r}_2,t_2)=\langle|A_1+A_2|^2\rangle,
\end{eqnarray}
where $(\mathbf{r}_j,t_j)$ is the space-time coordinates for the photon detection event at D$_j$ ($j=1$ and 2). $\langle...\rangle$ is ensemble average by taking all the possible realizations into account. $A_1$ and $A_2$ are the corresponding probability amplitudes for the above two alternatives, respectively. Based on the calculations in \cite{shih-book} and Appendix A of this paper, the value of $g^{(2)}(0)$ can be approximately estimated by the ratio between the  number of total terms and the number of autocorrelation terms in Eq. (\ref{g2-one}). There are 4 terms after the modulus square is calculated and 2 autocorrelation terms. The normalized second-order coherence function, $g^{(2)}(0)$, equals 2 in this case, which is consistent with the conclusions in \cite{brown-1957,brown-1958,purcell,fano}.

When there are two RGs, there are four different alternatives for two photons to trigger a two-photon coincidence count in the scheme in Fig. \ref{1}. These four alternatives are a$_1$-a$_2$-D$_1$ and b$_1$-b$_2$-D$_2$, a$_1$-a$_2$-D$_2$ and b$_1$-b$_2$-D$_1$,  a$_1$-b$_2$-D$_2$ and b$_1$-a$_2$-D$_1$, a$_1$-b$_2$-D$_1$ and b$_1$-a$_2$-D$_2$, respectively. a$_1$-a$_2$-D$_1$ means photon at a$_1$ goes to a$_2$ and then detected by D$_1$. The meanings of other symbols are similar. If these four different alternatives are indistinguishable, the second-order coherence function with two RGs is \cite{feynman-qed,feynman-path}
\begin{eqnarray}\label{g2-two}
G^{(2)}_2(\mathbf{r}_1,t_1;\mathbf{r}_2,t_2)=\langle|A_1+A_2+A_3+A_4|^2\rangle,
\end{eqnarray}
where $A_1$, $A_2$, $A_3$, and $A_4$ are the corresponding probability amplitudes for the above four alternatives, respectively. The number of total terms after modulus square in Eq. (\ref{g2-two}) is 4$^2$. The number of autocorrelation terms is 4. Hence  $g^{(2)}(0)$ equals $4^2/4$ for two RGs in Fig. \ref{1}, in which superbunching is expected.

\begin{figure}[htb]
\centering
\includegraphics[width=65mm]{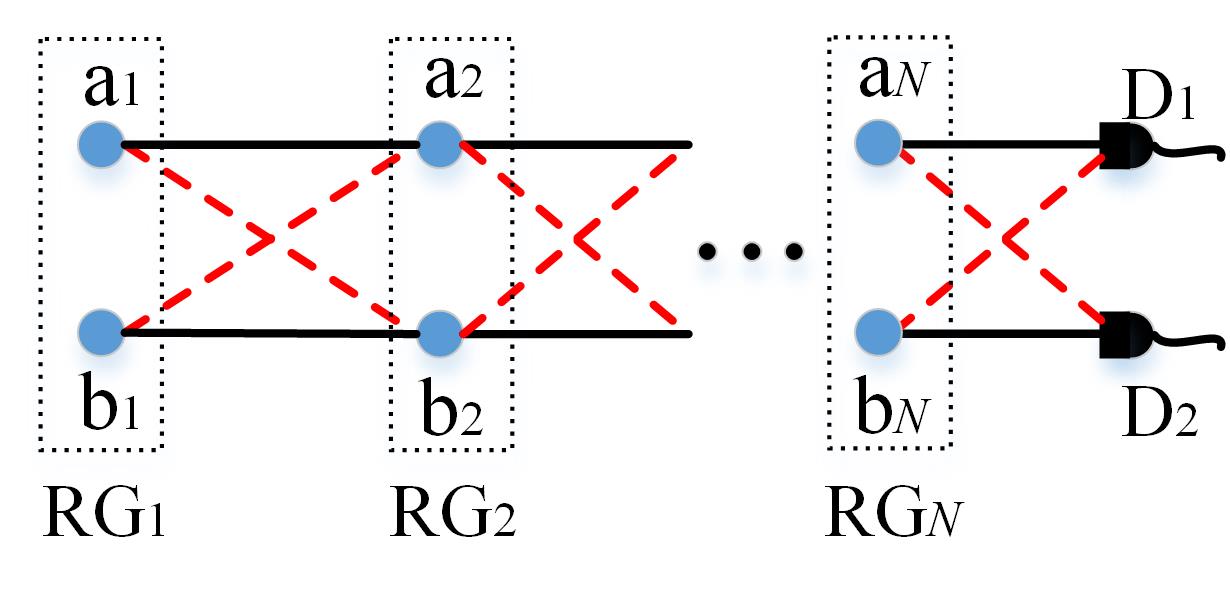}%
\caption{(Color online) Different pathes for two photons to trigger a two-photon coincidence count in superbunching scheme. a$_j$ and b$_j$ are the possible positions of photons on RG$_j$ ($j=1$, 2,..., $N$). D$_1$ and D$_2$ are two single-photon detectors in a HBT interferometer. The outputs of these two detectors are sent to a two-photon coincidence counting system to measure the two-photon coincidence count, which is not shown in the figure. }\label{2}
\end{figure}

The same method can be employed to estimate the normalized second-order coherence function of $N$ RGs in Fig. \ref{1}. There are 2$^N$ different alternatives for two photons to trigger a two-photon coincidence count for $N$ RGs in Fig. \ref{1}, which can be understood in the following way. There are $2^1$ and $2^2$ different alternatives for one and two RGs in Fig. \ref{1}, respectively. We can assume that there are $2^{N-1}$ different alternatives to trigger a two-photon coincidence count for $N-1$ RGs in Fig. \ref{1}. By adding the $N$th RG, there will be two more possible positions, a$_{N}$ and b$_{N}$, for the photons as shown in Fig. \ref{2}. There are $2^{N-1}$ different alternatives for the photon at a$_N$ goes to D$_1$ and the photon at b$_N$ goes to D$_2$. When exchanging the orders, \textit{i.e.}, the photon at a$_N$ goes to D$_2$ and the photon at b$_N$ goes to D$_1$, there are $2^{N-1}$ different alternatives, too. Hence the total number of alternatives to trigger a two-photon coincidence count for $N$ RGs in Fig. \ref{1} is $2^{N-1}$+$2^{N-1}$, which equals $2^{N}$.

If all the $2^N$ different alternatives are indistinguishable, the second-order coherence function for $N$ RGs is
\begin{eqnarray}\label{g2n}
G^{(2)}_N(\mathbf{r}_1,t_1;\mathbf{r}_2,t_2)=\langle|\sum^{2^N}_{j=1} A_j|^2\rangle,
\end{eqnarray}
where $A_j$ is the $j$th probability amplitude for the photons at a$_1$ and b$_1$ goes to D$_1$ and D$_2$ to trigger a two-photon coincidence count. The number of total terms in Eq. (\ref{g2n}) after modulus square is $(2^{N})^2$ and  the number of autocorrelation terms is $2^{N}$. The normalized second-order coherence function of $N$ RGs in the scheme shown in Fig. \ref{1} is
\begin{eqnarray}\label{g20}
g^{(2)}_N(0)=\frac{(2^N)^2}{2^N}=2^N,
\end{eqnarray}
where superbunching is expected for $N$ $(N\geq 2)$ RGs.

\subsection{The second-order temporal coherence function of superbunching pseudothermal light}

In this section, we will calculate the second-order temporal coherence functions for one and two RGs in the superbunching pseudothermal light scheme, respectively.

The Feynman's photon propagator for a point light source is \cite{QFT}
\begin{equation}\label{green}
K_{\alpha\beta}=\frac{\text{exp}[-i(\omega_{\alpha}
t_{\alpha\beta}-\mathbf{k}_{\alpha\beta}\cdot
\mathbf{r}_{\alpha\beta})]}{r_{\alpha\beta}},
\end{equation}
which is the same as the Green function for a point light source in classical optics \cite{born}. $\mathbf{r}_{\alpha\beta}$ equals $\mathbf{r}_{\beta}-\mathbf{r}_{\alpha}$, which is the position vector of the photon at  $\mathbf{r}_\alpha$ goes to $\mathbf{r}_\beta$. $\mathbf{r}_\alpha$ and $\mathbf{r}_\beta$ are two position vectors. $r_{\alpha\beta}$ is the distance between $\mathbf{r}_\alpha$ and $\mathbf{r}_\beta$, which equals $|\mathbf{r}_{\alpha\beta}|$. $\mathbf{k}_{\alpha\beta}$ and  $\omega_{\alpha}$ are the wave vector and frequency of the photon at  $\mathbf{r}_\alpha$ goes to $\mathbf{r}_\beta$, respectively.  $t_{\alpha\beta}$ equals $t_\beta-t_\alpha$, which is time for the photon at  $\mathbf{r}_\alpha$ goes to $\mathbf{r}_\beta$. $t_\alpha$ and $t_\beta$ are the time for the photon at $\mathbf{r}_\alpha$ and $\mathbf{r}_\beta$, respectively.

For simplicity, we will concentrate on the temporal correlation. The propagator in Eq. (\ref{green}) can be simplified as
\begin{equation}\label{green-t}
K_{\alpha\beta}\propto e^{-i\omega_{\alpha}(t_{\beta}-t_\alpha)}
\end{equation}
by ignoring the spatial part. The second-order coherence function in Eq. (\ref{g2-one}) can be rewritten as
\begin{eqnarray}\label{g2-one-t}
G^{(2)}_1(t_1,t_2)&=&\langle|e^{i\varphi_{a1}}K_{a1D1}e^{i\varphi_{b1}}K_{b1D2}\nonumber\\
&&+e^{i\varphi_{a1}}K_{a1D2}e^{i\varphi_{b1}}K_{b1D1}|^2\rangle,
\end{eqnarray}
where $\varphi_{a1}$ and $\varphi_{b1}$ are the initial phases of photons at a$_1$ and b$_1$, respectively. The initial phases of photons in thermal light are random \cite{loudon-book}.  $t_1$ and $t_2$ are short for t$_{D1}$ and t$_{D2}$, which are the time for photon detection events at D$_1$ and D$_2$, respectively.  Substituting Eq. (\ref{green-t}) into Eq. (\ref{g2-one-t}), it is straightforward to have
\begin{eqnarray}\label{g2-one-t1}
&&G^{(2)}_1(t_1-t_2) \propto 2+2\text{Re}[e^{-i\omega_{a1}(t_1-t_2)}e^{-i\omega_{b1}(t_1-t_2)}],
\end{eqnarray}
where Re is the real part of the complex expression. Assuming the frequency bandwidth of the light scattered by RG$_1$ is $\Delta \omega_1$, the normalized second-order temporal coherence function of one RG in Fig. \ref{1} is  \cite{shih-book,zhou-pra}
\begin{eqnarray}\label{g2-one-t1-2}
g^{(2)}_1(t_1-t_2) = 1+\text{sinc}^2\frac{\Delta\omega_{1}(t_1-t_2)}{2},
\end{eqnarray}
where sinc$(x)$ equals $sin(x)/x$. When the value of $|t_1-t_2|$ is large enough, $g^{(2)}_1(t_1-t_2)$ equals 1, which means the detections of these two photons are independent in this condition. $g^{(2)}_1(t_1-t_2)$ equals 2 when $t_1-t_2$ equals 0, which means photons in thermal light have tendency to come in bunches. This phenomenon is called two-photon bunching effect, which was first observed by Hanbury Brown and Twiss \cite{hbt,hbt-1}.

With the same method above, we can have the second-order temporal coherence function for two RGs in the scheme shown in Fig. \ref{1},
\begin{eqnarray}\label{g2-two-t1}
&&g^{(2)}_2(t_1-t_2)\\
&=&  [1+\text{sinc}^2\frac{\Delta\omega_{1}(t_1-t_2)}{2}][1+\text{sinc}^2\frac{\Delta\omega_{2}(t_1-t_2)}{2}],\nonumber
\end{eqnarray}
where the detail calculations can be found in Appendix A. $\Delta\omega_{1}$ and $\Delta\omega_{2}$ are the frequency bandwidths of thermal light at RG$_1$ and RG$_2$, respectively. When the value of $|t_1-t_2|$ is large enough, $g^{(2)}_2(t_1-t_2)$ equals 1, which means the detections of two photons are independent. When $t_1-t_2$ equals 0, $g^{(2)}_2(t_1-t_2)$ equals 4, which means superbunching can be observed.

\section{Experiments}\label{experiments}

\begin{figure}[htb]
\centering
\includegraphics[width=70mm]{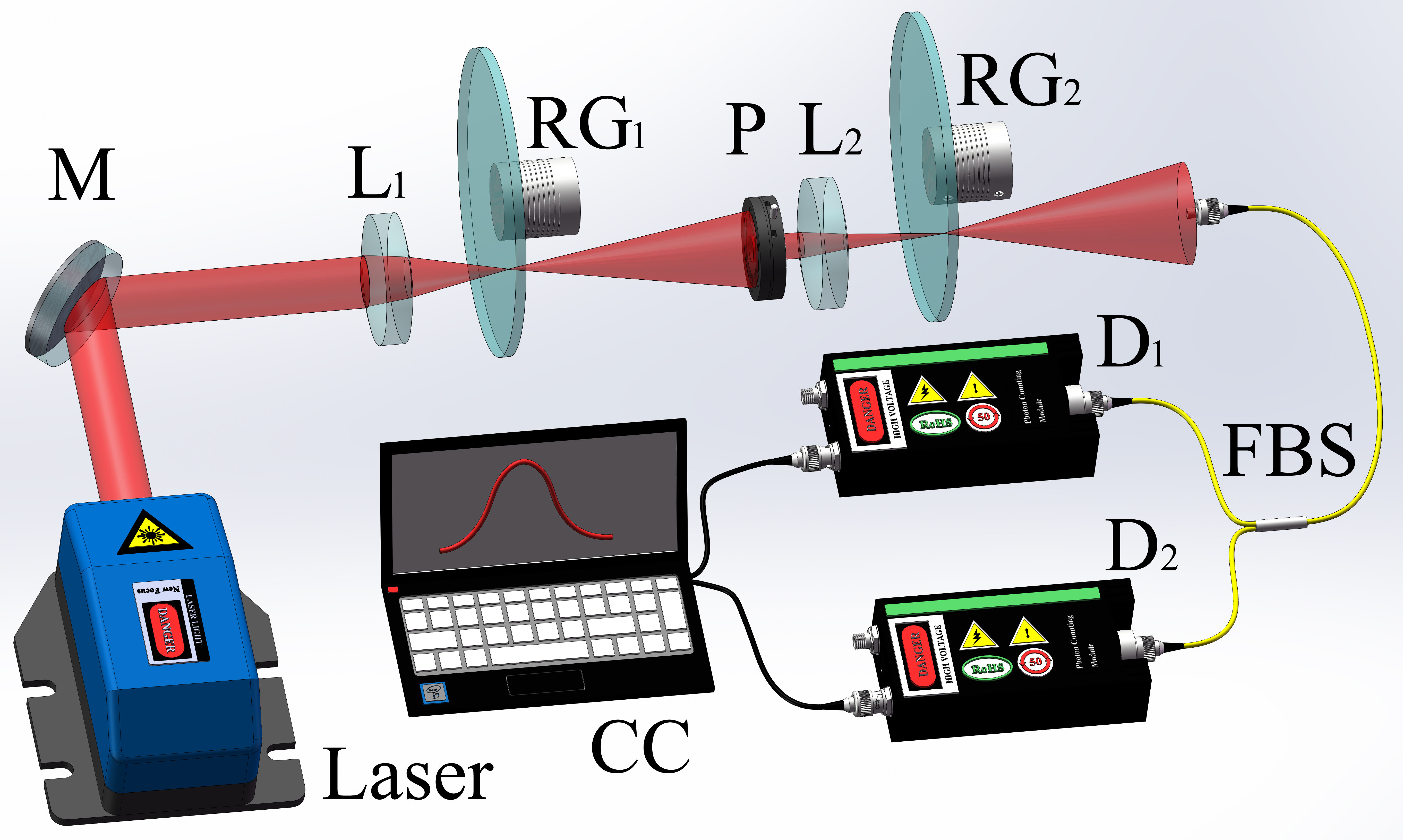}%
\caption{(Color online) Experimental setup for superbunching pseudothermal light source with two RGs. Laser: single-mode continuous-wave laser. M: Mirror. L: Lens. RG: Rotating groundglass. P: Pinhole. FBS: Non-polarized 50:50 fiber beam splitter. D: Single-photon detector. CC: Two-photon coincidence count detection system. }\label{3}
\end{figure}

The experimental setup to observe superbunching pseudothermal light with two RGs is shown in Fig. \ref{3}. The employed laser is a linearly-polarized single-mode continuous-wave laser with central wavelength at 780 nm and frequency bandwidth of 200 kHz (Newport, SWL-7513). M is a mirror. L$_1$ is a focus lens with focus length of 50 mm. RG$_1$ and RG$_2$ are two rotating groundglass. P is a pinhole. The distance between L$_1$ and RG$_1$ is 50 mm. The distance between RG$_1$ and the pinhole is 240 mm. The transverse coherence length of pseudothermal light generated by RG$_1$ is 4 mm in the pinhole plane.  The diameter of the pinhole is 1.2 mm, which is less than the coherence length. Only the light within one coherence area can pass the pinhole. The second lens with focus length of 25 mm, L$_2$, is employed to focus the light onto RG$_2$. The distance between L$_2$ and RG$_2$ is 28 mm, which is determined by minimizing the size of light spot on RG$_2$. The reason why the distance between L$_2$ and RG$_2$ is larger than the focus length of L$_2$ is that light scattered by RG$_1$ is diffuse instead of parallel. A non-polarized 50:50 fiber beam splitter (FBS) is employed to measure the second-order temporal coherence function. The distance between RG$_2$ and the collector of FBS is 700 mm.  The diameter of the collector of FBS is 5 $\mu$m, which is much less than the coherence length of pseudothermal light generated by RG$_2$ in the same plane ($\sim$13 mm). D$_1$ and D$_2$ are two single-photon detectors (PerkinElmer, SPCM-AQRH-14-FC). CC is two-photon coincidence count detection system (Becker \& Hickl GmbH, DPC-230).

\begin{figure}[htb]
\centering
\includegraphics[width=70mm]{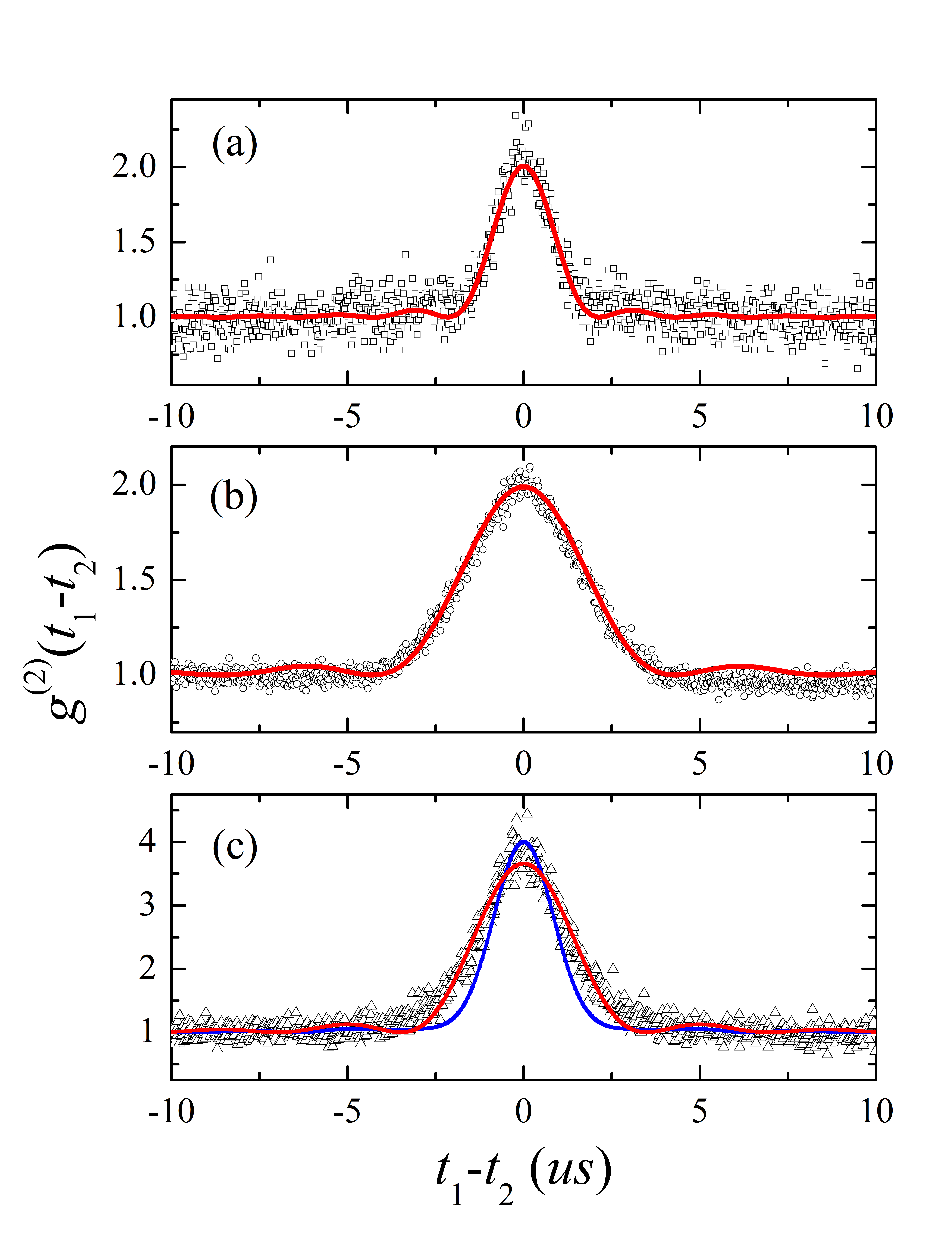}%
\caption{(Color online) Measured second-order temporal coherence functions. $g^{(2)}(t_1-t_2)$ is the normalized second-order coherence function. $t_1-t_2$ is the time difference between two single-photon detection events within a two-photon coincidence count. The squares, circles, and triangles are measured results. The red lines are theoretical fittings by employing Eq. (\ref{g2-one-t1-2}). (a) is measured when RG$_1$ is not rotating while RG$_2$ is rotating at 12 Hz. (b) is measured when RG$_1$ is rotating at 40 Hz while RG$_2$ is not rotating. (c) is measured when RG$_1$ and RG$_2$ are rotating at 40 Hz and 12 Hz, respectively. The blue line in (c) is the product of the red lines in (a) and (b).}\label{4}
\end{figure}

We first measure the second-order temporal coherence function of usual pseudothermal light \cite{martienssen}. Figure \ref{4}(a) shows the measured normalized second-order temporal coherence function when RG$_1$ is not rotating while RG$_2$ is rotating at 12 Hz. $g^{(2)}(t_1-t_2)$ is the normalized second-order coherence function and $t_1-t_2$ is the time difference between the two single-photon detection events within a two-photon coincidence count. The squares are the measured results, which are normalized according to the background. The red line is theoretical fitting by employing Eq. (\ref{g2-one-t1-2}). The measured coherence time and $g^{(2)}(0)$ of pseudothermal light in Fig. \ref{4}(a) are $1.08 \pm 0.01$ $\mu $s and $2.01 \pm 0.02$, respectively. Figure \ref{4}(b) shows the measured results when RG$_1$ is rotating at 40 Hz while RG$_2$ is not rotating. The circles are measured results and the red line is the fitting of the measured results by employing Eq. (\ref{g2-one-t1-2}). The measured coherence time and $g^{(2)}(0)$ of pseudothermal light in Fig. \ref{4}(b) are $2.15 \pm 0.03$ $\mu $s and $1.99 \pm 0.01$, respectively. Figure \ref{4}(c) is the measured second-order coherence function when RG$_1$ and RG$_2$ are rotating with speed of 40 and 12 Hz, respectively. The triangles are measured results and the red line is theoretical fitting of the data by employing Eq. (\ref{g2-one-t1-2}). The measured coherence time in Fig. \ref{4}(c) is $1.74 \pm 0.02$ $\mu $s.  The ratio between the peak and the background in Fig. \ref{4}(c) is much larger than the ones in Figs. \ref{4}(a) and \ref{4}(b). The normalized second-order coherence function, $g^{(2)}(0)$, equals $3.66\pm0.02$, in which superbunching is observed.

The blue line in Fig. \ref{4}(c) is the product of the two fitted lines in Figs. \ref{4}(a) and \ref{4}(b). It is consistent with the measured results except the calculated line is narrower. The reason may be the conditions to measure Fig. \ref{4}(c) are not exactly the same as the ones to measure Fig. \ref{4}(a). When measuring the temporal coherence function in  Fig. \ref{4}(a), we manually rotate RG$_1$ to a certain position to ensure that the single-photon counting rates of both detectors are at 5000 c/s level. The size of light spot on RG$_2$ did not vary during the whole measurement. However, RG$_1$ is rotating during the measurement of coherence function in Fig. \ref{4}(c), in which the size of light spot on RG$_2$ varies during the whole measurement. The difference between these two measurements may cause the deviation between the blue line and the measured results in Fig. \ref{4}(c). However, two-photon superbunching effect is observed in Fig. \ref{4}(c) from no matter the red line or the blue line, which means that the principle of superbunching pseudothermal light source in Fig. \ref{1} works.

\section{Discussions}\label{discussions}

\subsection{Why superbunching can be observed in our scheme}

In the last two sections, we have employed two-photon interference theory to predict that superbunching can be observed in the scheme shown in Fig. \ref{1} and experimentally confirmed it. The key to observed superbunching in the scheme is that all the different alternatives to trigger a two-photon coincidence count are in principle indistinguishable when there are more than one RGs. If these different alternatives to trigger a two-photon coincidence count are distinguishable, the second-order coherence function is \cite{feynman-path}
\begin{eqnarray}\label{g2n-dis}
G^{(2)}_N(\mathbf{r}_1,t_1;\mathbf{r}_2,t_2)=\langle\sum^{2^N}_{j=1}| A_j|^2\rangle,
\end{eqnarray}
where the probabilities instead of probability amplitudes are summed. There is no cross terms in Eq. (\ref{g2n-dis}). The ratio between the number of total terms and the number of autocorrelation terms is 1, which means $g^{(2)}_N(0)$ equals 1. No superbunching can be observed if all the alternatives are distinguishable.

In the scheme shown in Fig. \ref{1}, the necessary and sufficient condition for these different alternatives are indistinguishable is the photons are indistinguishable \cite{liu-2015-oc}. Photons are indistinguishable if they are within the same coherence volume \cite{mandel-book,martienssen}. Coherence volume is equal to the product of transverse coherence area and longitudinal coherence length. If a pinhole with diameter less than the transverse coherence length of light is employed to filter out the light, the photons passing through the pinhole are within the same coherence area. All the photons within the coherence time are indistinguishable in this case. This is what we have done in the scheme in Fig. \ref{1} and in the experiment in Fig. \ref{3}. A pinhole between  RG$_j$ and RG$_{j+1}$ is employed to filter out the photons within one coherence area of pseudothermal light generated by RG$_j$ ($j=1$, 2, ..., $N$). All the different alternatives are indistinguishable if a pinhole is employed to filter out photons before each RG.

In the early work by Hong \textit{et al.} \cite{hong}, superbunching is also observed by adding more alternatives via a modified Michelson interferometer. They experimentally observed $g^{(2)}(0)$ equals $2.4 \pm 0.1$ and theoretically proved that $g^{(2)}(0)$ will increase to $2\times 1.5^N$ if $N$ interferometers were inserted in their scheme. However, it is very difficult to insert more than one modified michelson interferometers in their scheme. We have observed $g^{(2)}(0)$ equals $3.66 \pm 0.02$ by employing two RGs and one pinhole, which is much simpler than the scheme employed in \cite{hong}. Further more, the value of $g^{(2)}(0)$ will increase to $2^N$ if $N$ RGs were employed in our scheme. Comparing to the scheme by Hong \textit{et al.} \cite{hong}, our superbunching pseudothermal light source is much simpler and the value of $g^{(2)}(0)$ increases faster as more alternatives were added in the scheme.

\subsection{Revised scheme and classical interpretation}

The superbunching pseudothermal light source in Fig. \ref{1} can also be understood in classical theory \cite{glauber,glauber1,sudarshan}. The intensity of light after one RG obeys negative exponential distribution \cite{goodman-book},
\begin{eqnarray}\label{g2e}
P(I)=\frac{1}{\langle I \rangle }\text{exp}(-\frac{I}{\langle I \rangle }),
\end{eqnarray}
where $\langle I \rangle$ is the average intensity of the scattered light. If a pinhole is employed to filter out the light within one coherence area, the intensity of light after the pinhole will obey Eq. (\ref{g2e}). Even though the intensity of the filtered light is not constant, it is coherent since it is within one coherence area \cite{martienssen}. This light beam can be incident to anther RG to generate pseudothermal light. Based on the results in Appendix B, the second-order moments of the light intensity after $n$ ($n$=2, 3, 4, and 5) RGs is
\begin{eqnarray}\label{g2cq3-2}
\langle I^2 \rangle =\langle I \rangle^2 2^n.
\end{eqnarray}
The normalized second-order coherence function is \cite{mandel-book}
\begin{eqnarray}\label{g20-c}
g^{(2)}_n(0)\equiv \frac{\langle I^2 \rangle}{\langle I \rangle^2} = 2^n.
\end{eqnarray}
We did not prove this equation for any value of $n$. However, it is confirmed that this equation is true for $n$ equals 2, 3, 4, and 5. The result in Eq. (\ref{g20-c}) is consistent with the one in Eq. (\ref{g20}).

\begin{figure}[htb]
\centering
\includegraphics[width=48mm]{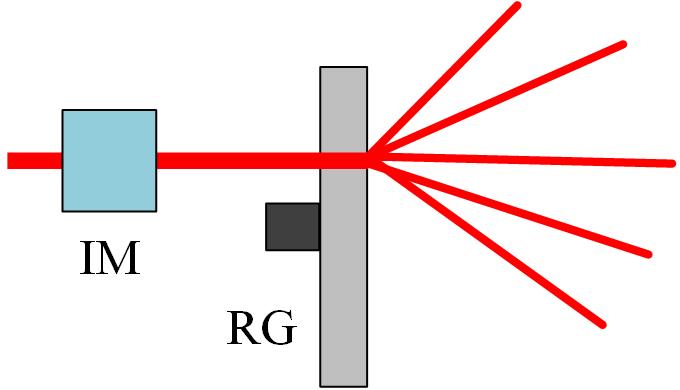}%
\caption{(Color online) Revised superbunching pseudothermal light scheme. The intensity modulator (IM) before rotating groundglass (RG) is used to modulate the intensity of the incident light without randomizing the phase.}\label{5}
\end{figure}

Most of the energy in the scheme shown in Fig. \ref{1} is wasted by scattering. From the classical point of view, all the RGs before the last one are employed to introduce certain intensity distribution. An intensity modulator (IM) can be employed to modulate the light intensity to obey the same distribution as the one of multiple RGs. Superbunching can also be observed in the scheme shown in Fig. \ref{5} if suitable intensity modulation is applied by IM.

There is no multiple alternatives before RG to trigger a two-photon coincidence count in the scheme shown in Fig. \ref{5}. How to understand the superbunching effect can also be observed in Fig. \ref{5} as the one in Fig. \ref{1}? For simplicity, let us take two RGs for example to explain the physics of these two schemes. In the scheme shown in Fig. \ref{1}, the light incident on RG$_2$ is filtered out by a pinhole from the pseudothermal light generated by RG$_1$. The intensity before RG$_2$ obey negative exponential distribution \cite{goodman-book}. This phenomenon can be understood by two-photon interference since there are different and indistinguishable alternatives to trigger a two-photon coincidence count. However, from classical point of view, the filtered light before RG$_2$ in Fig. \ref{1} can be mimicked by a light beam with the same negative exponential distribution as the one scattered by RG$_1$. There is no difference for RG$_2$ whether the incident light is filtered out by a pinhole in pseudothermal light or directly modulated by an IM as long as the intensities of light beams obey the same distribution and the light beams are coherent. The discussions can be generalized to the $N$ RGs case.

\section{Conclusions}\label{conclusion}

In summary, we have proposed a superbunching pseudothermal light source based on simple laboratory instruments such as lens, RG, and pinhole \textit{etc.}. Two-photon interference theory is employed to interpret the superbunching effect and it is found that the key to observe superbunching in our scheme is that all the different alternatives to trigger a two-photon coincidence count are in principle indistinguishable. $g^{(2)}(0)$ equals $3.66 \pm 0.02$ is observed with two RGs in the superbunching scheme and it is predicted that $g^{(2)}(0)$ can reach $2^N$ if $N$ RGs were employed. Based on the conclusions in classical theory, we suggested a different but equivalent superbunching pseudothermal light scheme by replacing all the RGs before the last one with an intensity modulator (IM). The revised scheme can be employed to observe superbunching as long as the intensity of the modulated light obey some certain distribution. Light intensity obeying negative exponential distribution and related distributions is discussed in this paper. It would be interesting to study whether superbunching can be observed or not when the intensity obeys other type of distributions.

The observed superbunching is in the temporal part, which is helpful to improve the visibility of temporal ghost imaging with classical light \cite{gi-t1}. Whether the spatial superbunching can be realized by analogy of the temporal superbunching is an interesting topic, too. The discussions of superbunching, in both the quantum and classical theories, are helpful to understand the physics of two-photon superbunching. This novel and simple superbunching pseudothermal light source will be an important tool to study thermal light ghost imaging, the   second- and higher-order interference of thermal light, and other possible applications of thermal light.

\section*{Acknowledgement}

This project is supported by National Science Foundation of China (No.11404255), Doctoral Fund of Ministry of Education of China (No.20130201120013), the 111 Project of China (No.B14040) and the Fundamental Research Funds for the Central Universities.

\section*{Appendix A: The second-order temporal coherence function of two RGs}\label{appendix-a}
\def\theequation{$A-$\arabic{equation}}
\setcounter{equation}{0}
\def\thefigure{$A-$\arabic{figure}}
\setcounter{figure}{0}

There are four different alternatives for two photons to trigger a two-photon coincidence count when two RGs are in the scheme shown in Fig. \ref{1}. The first one is a$_1$-a$_2$-D$_1$ and b$_1$-b$_2$-D$_2$, which means the photon at a$_1$ goes to a$_2$ and then detected by D$_1$ and the photon at b$_1$ goes to b$_2$ and then detected by D$_2$. The other three alternatives are a$_1$-a$_2$-D$_2$ and b$_1$-b$_2$-D$_1$, a$_1$-b$_2$-D$_2$ and b$_1$-a$_2$-D$_1$, and a$_1$-b$_2$-D$_1$ and b$_1$-a$_2$-D$_2$, respectively. If these four different alternatives are indistinguishable, the second-order coherence function of two RGs in the scheme shown in Fig. \ref{1} is
\begin{eqnarray}\label{g2-two-a1}
&&G^{(2)}_2(\mathbf{r}_1,t_1;\mathbf{r}_2,t_2)\nonumber\\
&=&\langle|A_{a1a2D1}A_{b1b2D2}+A_{a1a2D2}A_{b1b2D1}\nonumber\\
&&+A_{a1b2D2}A_{b1a2D1}+A_{a1b2D1}A_{b1a2D2}|^2\rangle.
\end{eqnarray}

With the same method as the one for one RG, we only consider the temporal part. There will be 16 terms after modulus square is evaluated in Eq. (\ref{g2-two-a1}). The 4 autocorrelation terms only contribute to the background. There are 12 cross-correlation terms left, which can be categorized into three groups. We can have the result of one group by calculating one term from the same group.

The first term need to be calculated is $A_{a1a2D1}A_{b1b2D2}A^*_{a1a2D2}A^*_{b1b2D1}$, where $A^*_{a1a2D2}$ is the complex conjugate of $A_{a1a2D2}$. The probability amplitude of two successive and independent event equals the product of these two different probability amplitudes \cite{feynman-path}. $A_{a1a2D1}$ can be written as $A_{a1a2}A_{a2D1}$. Other terms can be simplified in the same way. Substituting this relation and Eq. (\ref{green-t}) into the term above, we have
\begin{eqnarray}\label{g2-two-a2}
&&A_{a1a2D1}A_{b1b2D2}A^*_{a1a2D2}A^*_{b1b2D1} \nonumber\\
&=&e^{-i\omega_{a1}(t_{a2}-t_{a1})}e^{-i\omega_{a2}(t_{1}-t_{a2})}e^{-i\omega_{b1}(t_{b2}-t_{b1})}e^{-i\omega_{b2}(t_{2}-t_{b2})}\nonumber\\
&&\times e^{i\omega_{a1}(t_{a2}-t_{a1})}e^{i\omega_{a2}(t_{2}-t_{a2})}e^{i\omega_{b1}(t_{b2}-t_{b1})}e^{i\omega_{b2}(t_{1}-t_{b2})}\nonumber\\
&=&e^{-i\omega_{a2}(t_{1}-t_{2})}e^{i\omega_{b2}(t_{1}-t_{2})}.
\end{eqnarray}
The last term on the righthand side of Eq. (\ref{g2-two-a2}) is similar as the last term of Eq. (\ref{g2-one-t1}). If the frequency bandwidth of thermal light scattered by RG$_2$ is $\Delta \omega_2$, $A_{a1a2D1}A_{b1b2D2}A^*_{a1a2D2}A^*_{b1b2D1}$ can be calculated as
\begin{eqnarray}\label{g2-two-a3}
&&A_{a1a2D1}A_{b1b2D2}A^*_{a1a2D2}A^*_{b1b2D1} \nonumber\\
&=& \int\int_{\omega_0-\frac{\Delta \omega_2}{2}}^{\omega_0+\frac{\Delta \omega_2}{2}} e^{-i\omega_{a2}(t_{1}-t_{2})}e^{i\omega_{b2}(t_{1}-t_{2})} d\omega_{a2}d\omega_{b2}\nonumber\\
&=&(\Delta \omega_2)^2 \text{sinc}^2 \frac{\Delta\omega_2(t_1-t_2)}{2}.
\end{eqnarray}
$\omega_0$ is the central frequency of light and  $ \text{sinc}(x)$ equals $\sin(x)/x$. Other three terms, $A^*_{a1a2D1}A^*_{b1b2D2}A_{a1a2D2}A_{b1b2D1}$, $A_{a1b2D1}A_{b1a2D2}A^*_{a1b2D2}A^*_{b1a2D1}$, and $A^*_{a1b2D1}A^*_{b1a2D2}A_{a1b2D2}A_{b1a2D1}$ in the same group have the same result as the one of Eq. (\ref{g2-two-a3}).

The second term need to be calculated is $A_{a1a2D1}A_{b1b2D2}A^*_{a1b2D2}A^*_{b1a2D1}$. With the same method above, we have
\begin{eqnarray}\label{g2-two-a4}
&&A_{a1a2D1}A_{b1b2D2}A^*_{a1b2D2}A^*_{b1a2D1} \nonumber\\
&=&e^{-i\omega_{a1}(t_{a2}-t_{a1})}e^{-i\omega_{a2}(t_{1}-t_{a2})}e^{-i\omega_{b1}(t_{b2}-t_{b1})}e^{-i\omega_{b2}(t_{2}-t_{b2})}\nonumber\\
&&\times e^{i\omega_{a1}(t_{b2}-t_{a1})}e^{i\omega_{b2}(t_{2}-t_{b2})}e^{i\omega_{b1}(t_{a2}-t_{b1})}e^{i\omega_{a2}(t_{1}-t_{a2})}\nonumber\\
&=&e^{-i\omega_{a1}(t_{a2}-t_{b2})}e^{i\omega_{b1}(t_{a2}-t_{b2})}.
\end{eqnarray}
Assuming the frequency bandwidth of thermal light scattered by RG$_1$ is $\Delta \omega_1$, $A_{a1a2D1}A_{b1b2D2}A^*_{a1b2D2}A^*_{b1a2D1}$ can be calculated as
\begin{eqnarray}\label{g2-two-a5}
&&A_{a1a2D1}A_{b1b2D2}A^*_{a1b2D2}A^*_{b1a2D1} \nonumber\\
&=& \int\int_{\omega_0-\frac{\Delta \omega_1}{2}}^{\omega_0+\frac{\Delta \omega_1}{2}}e^{-i\omega_{a1}(t_{a2}-t_{b2})}e^{i\omega_{b1}(t_{a2}-t_{b2})}d\omega_{a1}d\omega_{b1}\nonumber\\
&=&(\Delta \omega_1)^2 \text{sinc}^2 \frac{\Delta\omega_1(t_{a2}-t_{b2})}{2},
\end{eqnarray}
where the central frequency is assumed to be the same during scattering in different RGs. $t_{a2}$ is related to $t_1$ by the relation, $t_{a2}=t_1-r_{a2D1}/c$, where $r_{a2D1}$ is the distance between $\mathbf{r}_{a2}$ and $\mathbf{r}_{D1}$. $c$ is the velocity of light in the vacuum. In the similar way, $t_{b2}$ is related to $t_2$ by  $t_{b2}=t_2-r_{b2D2}/c$. Point light source and symmetrical positions for D$_1$ and D$_2$ are assumed in the calculations of temporal correlation. $r_{a2D1}$ equals $r_{b2D2}$. $t_{a2}-t_{b2}$ can be replaced by $t_1-t_2$ in Eq. (\ref{g2-two-a5}),
\begin{eqnarray}\label{g2-two-a6}
&&A_{a1a2D1}A_{b1b2D2}A^*_{a1b2D2}A^*_{b1a2D1} \nonumber\\
&=&(\Delta \omega_1)^2 \text{sinc}^2 \frac{\Delta\omega_1(t_{1}-t_{2})}{2}.
\end{eqnarray}
The terms, $A^*_{a1a2D1}A^*_{b1b2D2}A_{a1b2D2}A_{b1a2D1}$, $A_{a1a2D2}A_{b1b2D1}A^*_{a1b2D1}A^*_{b1a2D2}$, and $A^*_{a1a2D2}A^*_{b1b2D1}A_{a1b2D1}A_{b1a2D2}$ in the same group  have the same result as the one in Eq. (\ref{g2-two-a6}).

The third term needs to be calculated is $A_{a1a2D1}A_{b1b2D2}A^*_{a1b2D1}A^*_{b1a2D2}$, which is
\begin{eqnarray}\label{g2-two-a7}
&&A_{a1a2D1}A_{b1b2D2}A^*_{a1b2D1}A^*_{b1a2D2} \\
&=&e^{-i\omega_{a1}(t_{a2}-t_{a1})}e^{-i\omega_{a2}(t_{1}-t_{a2})}e^{-i\omega_{b1}(t_{b2}-t_{b1})}e^{-i\omega_{b2}(t_{2}-t_{b2})}\nonumber\\
&&\times e^{i\omega_{a1}(t_{b2}-t_{a1})}e^{i\omega_{b2}(t_{1}-t_{b2})}e^{i\omega_{b1}(t_{a2}-t_{b1})}e^{i\omega_{a2}(t_{2}-t_{a2})}.\nonumber\\
&=&e^{-i\omega_{a1}(t_{a2}-t_{b2})}e^{i\omega_{b1}(t_{a2}-t_{b2})}e^{-i\omega_{a2}(t_{1}-t_{2})}e^{i\omega_{b2}(t_{1}-t_{2})}.\nonumber
\end{eqnarray}
Integrating over the frequency bandwidths of thermal light scattered by RG$_1$ and RG$_2$, we have
\begin{eqnarray}\label{g2-two-a8}
&&A_{a1a2D1}A_{b1b2D2}A^*_{a1b2D1}A^*_{b1a2D2} \\
&=&(\Delta \omega_1 \Delta \omega_2)^2 \text{sinc}^2 \frac{\Delta\omega_1(t_{1}-t_{2})}{2}\text{sinc}^2 \frac{\Delta\omega_2(t_1-t_2)}{2}. \nonumber
\end{eqnarray}
The other three terms, $A^*_{a1a2D1}A^*_{b1b2D2}A_{a1b2D1}A_{b1a2D2} $, $A_{a1a2D2}A_{b1b2D1}A^*_{a1b2D2}A^*_{b1a2D1}$, and $A^*_{a1a2D2}A^*_{b1b2D1}A_{a1b2D2}A_{b1a2D1}$ in the same group  have the same result as the one in Eq. (\ref{g2-two-a8}).

In the calculations of Eqs. (\ref{g2-two-a3}) and (\ref{g2-two-a5}), we have ignored the integral of the constant, 1, for RG$_1$ and RG$_2$, respectively. If we take this factor into account and also integrate the autocorrelation terms, the second-order temporal coherence function with two RGs in the scheme in Fig. \ref{1} is
\begin{eqnarray}\label{g2-two-a9}
&&G^{(2)}_2(t_1-t_2)\nonumber\\
&\propto&4 (\Delta \omega_1 \Delta \omega_2)^2 [1+ \text{sinc}^2 \frac{\Delta\omega_1(t_{1}-t_{2})}{2}  \nonumber\\
&&+\text{sinc}^2 \frac{\Delta\omega_2(t_1-t_2)}{2}\rangle\nonumber\\
&&+\text{sinc}^2 \frac{\Delta\omega_1(t_{1}-t_{2})}{2}\text{sinc}^2 \frac{\Delta\omega_2(t_1-t_2)}{2}].
\end{eqnarray}
All the 16 terms in Eq. (\ref{g2-two-a1}) are calculated. Rearranging the terms on the righthand side of Eq. (\ref{g2-two-a9}), the normalized second-order temporal coherence function can be expressed as
\begin{eqnarray}\label{g2-two-10}
&&g^{(2)}_2(t_1-t_2)\\
&=&[1+ \text{sinc}^2 \frac{\Delta\omega_1(t_{1}-t_{2})}{2}][1+ \text{sinc}^2 \frac{\Delta\omega_2(t_{1}-t_{2})}{2}],\nonumber
\end{eqnarray}
in which Eq. (\ref{g2-two-t1}) is obtained.

\section*{Appendix B: Calculations of $g^{2}(0)$ in classical theory}\label{appendix-b}
\def\theequation{$B-$\arabic{equation}}
\setcounter{equation}{0}

We will follow the method given by Goodman to show how the second-order coherence function in the scheme shown in Fig. \ref{1} can be calculated in classical theory \cite{goodman-book}. The probability density function of pseudothermal light generated by scattering single-mode continuous-wave laser light on a rotating groundglass is negative exponential distribution \cite{martienssen}. If the intensity of the incident light varies, the conditional density function of the scattered light should be
\begin{eqnarray}\label{g2c1}
P_{I|x}(I|x)=\frac{1}{x}\text{exp}(-\frac{I}{x}),
\end{eqnarray}
where $x$ is proportional to the intensity of the incident light. If the incident light is filtered out as the one shown in Fig. \ref{1}, the intensity, $x$, obeys negative exponential distribution, too. The density distribution of the light intensity after RG$_2$ is
\begin{eqnarray}\label{g2c2}
P_{I}(I)=\int_{0}^{\infty}\frac{1}{x}\text{exp}(-\frac{I}{x})\cdot \frac{1}{\langle I \rangle }\text{exp}(-\frac{x}{\langle I \rangle })dx,
\end{eqnarray}
where $\langle I \rangle $ is the average intensity of the scattered light after RG$_2$. Equation (\ref{g2c2}) can be simplified as \cite{goodman-book}
\begin{eqnarray}\label{g2c3}
P_{I}(I)=\frac{2}{\langle I \rangle } K_0 (2\sqrt{\frac{I}{\langle I \rangle}}),
\end{eqnarray}
where $K_{0}(x)$ is the modified Bessel function of the second kind, order 0. The $q$th moments of the intensity is
\begin{eqnarray}\label{g2cq}
\langle I^q \rangle =\int_{0}^{\infty}I^q P_I(I)dI=\langle I \rangle^q (q!)^2.
\end{eqnarray}

In classical theory, the normalized second-order coherence function is defined as \cite{mandel-book}
\begin{eqnarray}\label{g2-def}
g^{(2)}(\mathbf{r}_1,t_1;\mathbf{r}_2,t_2)=\frac{\langle I(\mathbf{r}_1,t_1)I(\mathbf{r}_2,t_2) \rangle}{\langle I(\mathbf{r}_1,t_1)\rangle \langle I(\mathbf{r}_2,t_2) \rangle},
\end{eqnarray}
where $I(\mathbf{r}_j,t_j)$ is the intensity of light at space-time coordinate $(\mathbf{r}_j,t_j)$  ($j=1$ and 2). When these two detectors are at symmetrical positions, the normalized second-order coherence function can be simplified as,
\begin{eqnarray}\label{g2-def0}
g^{(2)}(0)=\frac{\langle I^2 \rangle }{\langle I \rangle^2 }.
\end{eqnarray}

Substituting Eq. (\ref{g2cq}) into Eq. (\ref{g2-def0}), the normalized second-order coherence function with two RGs in Fig. \ref{1} equals 4, which is consistent with the result of Eq. (\ref{g20}) in quantum theory.

With the same method, we can calculate the normalized second-order coherence function for more than two RGs. If there are three RGs, the input intensity of RG$_3$ is given by Eq. (\ref{g2c3}), the density function of the light intensity after RG$_3$ is given by
\begin{eqnarray}\label{g2c5}
P_{I}(I)=\int_{0}^{\infty}\frac{2}{x}K_0(2\sqrt{\frac{I}{x}})\cdot \frac{1}{\langle I \rangle }\text{exp}(-\frac{x}{\langle I \rangle })dx.
\end{eqnarray}
There is no analytical expression for Eq. (\ref{g2c5}). However, only the moment is needed for calculating the normalized second-order coherence function. With the help of Eqs. (\ref{g2cq}) and (\ref{g2c5}), the $q$th moment of the light intensity after three RGs is
\begin{eqnarray}\label{g2cq3}
\langle I^q \rangle =\langle I \rangle^q (q!)^3.
\end{eqnarray}
The corresponding normalized second-order coherence function, $g^{(2)}(0)$, equals 8. With the same method above, the $q$th moments of intensity after four and five RGs are $\langle I \rangle^q (q!)^4$ and $\langle I \rangle^q (q!)^5$, respectively, which correspond to the normalized second-order coherence functions equal 16 and 32, respectively. These results are consistent with the one in Eq. (\ref{g20}).

Employing classical theory to calculate the normalized second-order coherence function for more than five RGs is straightforward. However, the process may be cumbersome. The results above is sufficient to prove that if we can employ intensity modulator (IM) to modulate the intensity of light to obey negative exponential distribution before RG, $g^{(2)}(0)$ should equal 4. For the condition of more than two RGs, numerical method can be employed to calculate the intensity distribution and then apply the distribution on the IM. The experimental realization of superbunching pseudothermal light with $g^{(2)}(0)=2^N$ should be possible for $N$ larger than 2 in the scheme shown in Fig. \ref{5}.

\end{document}